\documentclass[showpacs,preprintnumbers,amsmath,amssymb,amsfonts]{revtex4}
\usepackage{graphicx}
\usepackage{dcolumn}
\usepackage{bm}
\newcommand{\trace}{\mathop{\rm Tr}\nolimits}
\newcommand{\conj}{\stackrel{?}{=}}
\newcommand{\coin}{\stackrel{?}{\geq}}
\newcommand{\colein}{\stackrel{?}{\leq}}
\newcommand{\braket}[2]{\langle #1 | #2 \rangle}
\newcommand{\beq}{\begin{equation}}
\newcommand{\eeq}{\end{equation}}
\newcommand{\bea}{\begin{eqnarray}}
\newcommand{\eea}{\end{eqnarray}}
\newcommand{\ket}[1]{| #1 \rangle}
\newcommand{\bra}[1]{\langle #1 |}
\newcommand{\proj}[1]{\ket{#1}\! \bra{#1}}

\renewcommand{\paragraph}{\section}

\newcommand{\qed}{\hfill$\square$\par\vskip12pt}

\begin{document}



\title{Strong superadditivity of the entanglement of formation
follows from its additivity}
\author{A.A. Pomeransky}
\email{pomeransky@irsamc.ups-tlse.fr}
\affiliation{ Laboratoire de Physique Th\'eorique,Universit\'e Paul Sabatier,
118 Route de Narbonne, 31062 Toulouse Cedex, France}
\begin{abstract}
The additivity of both the entanglement of formation and the
classical channel capacity is known to be a consequence of the
strong superadditivity conjecture. We show that, conversely,
the strong superadditivity conjecture follows from the 
additivity of the entanglement of formation; this means that
the two conjectures are equivalent and that the additivity of the
classical channel capacity is a consequence of them.
\end{abstract}
\maketitle
          


\section{Introduction}
Entanglement, an exclusively quantum property, makes possible numerous
new promising applications of quantum mechanics in computing, communication
and cryptography. One says that there is entanglement between different parts
of a quantum system if the state of the system cannot be represented
as a product of states or a statistical mixture of products. One of the
basic tasks of quantum information theory is to define the appropriate
quantitative characteristics of {\it how much} a state is entangled.
A simple and universal measure exists only for the case of 
two subsystems in a pure state. For a pure state $\psi$ of a system composed of the
subsystems $A$ and $B$ the entanglement $E(\psi)$ is given by the entropy
of the reduced density matrix:
\beq
E(\psi)=S(\trace_B(\proj{\psi}))=S(\trace_A(\proj{\psi})),
\eeq
where $S$ is the von Neumann entropy: $S(\rho)=-\trace\rho\log_2\rho$. 
Here and below the symbol $\trace$ with subscripts means the partial
trace over the corresponding subsystem (the subsystem $B$ or $A$ in this
case) and we consider only systems with finite-dimensional Hilbert
spaces. In contrast to the pure state case, for mixed states the different 
aspects of entanglement are characterized by different measures. For example,
the {\it entanglement cost}, the quantity of entanglement required to
prepare a given state, in general differs from the {\it distillable entanglement},
the quantity of entanglement which can be extracted from a given state.
One of the most important and widely used measures is the {\it entanglement
of formation} (EoF). It was introduced in \cite{bennett} as the least expected 
entanglement of any ensemble of pure states realizing $\rho$:
\beq\label{edef}
E_F(\rho) = \min_{\{p_i,\psi_i\}} \sum_i p_i E(\psi_i),
\eeq
where an ensemble $\{p_i,\psi_i\}$ realizes $\rho$ if $\rho=\sum_i p_i \proj{\psi_i}$,
that is if a pure state $\psi_i$ can be found in $\rho$ with probability $p_i$.
We will call {\it optimal ensemble} of $\rho$  an ensemble for which the minimum 
is attained.

The physical interpretation of the EoF depends on whether it is additive or not.
Given two states $\rho_1$ and $\rho_2$ of two separate systems $1$ and $2$
(each being a bipartite system with the parts $1A$, $1B$ and $2A$, $2B$ 
respectively, we always consider entanglement between $A$ and $B$), what is 
the EoF of the state $\rho_1\otimes\rho_2$ of the composite system? It has been
conjectured, that the EoF is additive: 
\beq
E_F(\rho_1 \otimes \rho_2)\conj\,E_F(\rho_1)+E_F(\rho_2),
\eeq
the EoF of the composite system is the sum of the EoF's of its parts. 
This conjecture has been proved for some particular classes of states.
Moreover the conjecture is supported by a number of numerical calculations
and no counterexamples has been found. It is known \cite{hayden}, that the 
entanglement cost $E_C$ of a state $\rho$ is equal to the asymptotic ratio 
of the EoF of $n$ copies of the state $\rho$ to the number of copies $n$,
that is $E_C=\lim_{n\rightarrow\infty} E_F(\rho^{\otimes n})/n$.
If the additivity conjecture is true, then the EoF gives us the entanglement cost
($E_C=E_F$), which would greatly simplify the problem of the practical calculation 
of $E_C$. It is natural to consider the more general problem of comparing the EoF 
of a system with the sum of the EoF's of its subsystems, and it is 
conjectured \cite{vw,msw} that the former is not less than the latter:
\beq\label{ssa}
E_F(\rho)\coin E_F(\trace_2\rho)+E_F(\trace_1\rho).
\eeq
This property is called strong superadditivity and it is not only
interesting on its own, but also because it implies the additivity of the 
EoF \cite{vw,msw}. It implies the additivity of the Holevo-Schumacher-Westmorland 
classical capacity of a quantum channel \cite{msw} too. The problem of the
additivity of this quantity is of considerable importance for the quantum 
communication theory \cite{ahw}, but it remains unresolved in the general case, 
though the additivity was proved for some particular classes of quantum channels.  
In this paper we uncover an even closer connection between the strong
superadditivity of the EoF, the additivity of the EoF and the additivity of the
classical channel capacity: we show that the additivity of the EoF {\it implies
the strong superadditivity} and therefore that these two conjectures are
equivalent and they imply the additivity of the classical channel capacity. 

The rest of the paper is organized as follows: in section 2 we discuss 
the convexity property of the EoF and introduce the crucial notion of
conjugate function. This and the majority of the other tools we use
were introduced in \cite{ab}. We also state in this section some related 
known facts. In section 3 we derive an equation which determines the optimal 
vectors for the conjugate function. This equation is used in section 4 to 
prove that the strong superadditivity of the EoF is equivalent to additivity.
We conclude with a discussion of the possible implications and with some
historical remarks.

\section{Convexity and the conjugate function}
One of the most important properties of the EoF is its convexity. 
Convexity means that for any set of density matrices $\rho_i$ and 
probabilities $p_i$ the EoF of the average density matrix $\rho=\sum_i p_i \rho_i$ 
is not greater than the average EoF:
\beq\label{conv}
E_F(\rho) \leq \sum_i p_i E_F(\rho_i).
\eeq
To convince oneself that the convexity holds one can consider the state $\rho$ as 
resulting from taking with probability $p_i$ an index $i$ and then preparing 
the system in a pure state from an optimal ensemble of $\rho_i$ with the 
probability corresponding to that pure state in the ensemble. The expected entanglement 
for the resulting pure state ensemble is equal to the r.h.s. of Eq. (\ref{conv}), 
and the least expected entanglement $E_F(\rho)$ is not greater than that.

Let us introduce following \cite{ab} an indispensable notion of the
conjugate function of the EoF. The transition from a function to its 
conjugate is a standard operation in convex analysis \cite{rock},
and for the EoF we obtain the following function of a Hermitian matrix $H$:
\beq\label{ddef}
E^*(H)=\max_{\rho}(\trace(\rho H)-E_F(\rho)),
\eeq
where maximization is performed over all density matrices $\rho$. 
Instead, one can maximize over pure states only \cite{ab}
\beq\label{dpur}
E^*(H)=\max_{\psi}(\bra{\psi}H\ket{\psi}-E(\psi)),
\eeq
because the expression $E_F(\rho)$ is the expected entanglement for an 
ensemble of pure states $\psi_i$, and the whole r.h.s. of Eq. (\ref{ddef}) 
is therefore also an average: 
\beq
\trace(\rho H)-E_F(\rho)=\sum_i p_i (\bra{\psi_i}H\ket{\psi_i}-E(\psi_i)),
\eeq
and an average cannot be greater than all averaging numbers.
Applying to a convex function the conjugation operation 
twice leaves it unchanged \cite{ab,rock}. For the EoF it means that
\beq\label{dd}
E_F(\rho)=\max_{H}(\trace(\rho H)-E^*(H)).
\eeq

Let us recall the statements of the strong superadditivity and additivity conjectures.
Consider a system, composed of two bipartite subsystems 1 and 2: the subsystem 1 consists
of parts $1A$ and $1B$, and the subsystem 2 consists of parts $2A$ and $2B$.
We always consider the entanglement between the subsystems $A$ and $B$.
The following conjectured property is called strong superadditivity:
\beq
E_F(\rho)\coin E_F(\rho_1)+E_F(\rho_2),
\eeq
where $\rho_{1,2}=\trace_{2,1}(\rho)$.
Additivity is another conjectured property:
\beq\label{add}
E_F(\rho_1 \otimes \rho_2)\conj\,E_F(\rho_1)+E_F(\rho_2).
\eeq
Let us note that additivity holds when the states $\rho_1$ and $\rho_2$ are pure.
One of the reasons why the strong superadditivity conjecture is interesting,
is that it implies the additivity of the EoF \cite{dvc,msw}. This is easy to see if 
we consider an optimal decomposition for $\rho_1$:
\bea
\rho_{1}=\sum_{i}p_i^{(1)}\proj{\psi_i^{(1)}},\;
E_F(\rho_{1})=\sum_{i}p_i^{(1)}E(\psi_i^{(1)})
\eea
and an analogous optimal decomposition for $\rho_2$.
We have a decomposition of the tensor product 
\beq
\rho_1 \otimes \rho_2=\sum_{ij}p_i^{(1)} p_j^{(2)}
\ket{\psi_i^{(1)}}\ket{\psi_j^{(2)}}\bra{\psi_j^{(2)}}\bra{\psi_i^{(1)}}.
\eeq
The mean EoF of this decomposition cannot exceed $E_F(\rho_1\otimes\rho_2)$:
\beq\label{sub}
E_F(\rho_1 \otimes \rho_2) \leq \sum_{ij}p_i^{(1)} p_j^{(2)}
E(\ket{\psi_i^{(1)}}\ket{\psi_j^{(2)}})=E_F(\rho_1)+E_F(\rho_2),
\eeq
where we used the additivity of the EoF for pure states.
The inequalities Eqs. (\ref{ssa}) and (\ref{sub}) combined give Eq. (\ref{add}).

Eq. (\ref{ssa}) holds for all states $\rho$ if and only if it
holds for pure states, that is if for all pure states $\psi$ \cite{dvc}:
\beq\label{pure}
E(\psi)\coin E_F(\trace_1(\proj{\psi}))+E_F(\trace_2(\proj{\psi})).
\eeq 
Indeed, let us consider an optimal decomposition of $\rho$:
\beq
\rho=\sum_i p_i\proj{\psi_i},\;\; E_F(\rho)= \sum_i p_iE(\psi_i).
\eeq
If Eq. (\ref{pure}) holds for these pure states $\psi_i$ then
\beq
E_F(\rho)= \sum_i p_i E(\psi_i)
\geq \sum_i p_i (E_F(\trace_2(\proj{\psi_i}))+E_F(\trace_1(\proj{\psi_i})))\nonumber
\eeq
Using the linearity of the trace: $\sum_i p_i \trace_1(\proj{\psi_i})=\trace_1\rho$ 
and the same for the subsystem 2, and using the convexity of the EoF Eq. 
(\ref{conv}), we obtain Eq. (\ref{ssa}) for the state $\rho$.

The strong superadditivity conjecture can be restated in terms of the conjugate
function $E^*(H)$. For this purpose, let us substitute Eq. (\ref{dd}) in the
r.h.s. of Eq. (\ref{pure}):
\begin{widetext}
\bea
E(\psi)&\coin& \max_{H_1} (\trace_1(\trace_2(\proj{\psi})H_1)-\,E^*(H_1))
+\max_{H_2} (\trace_2(\trace_1(\proj{\psi})H_2)-\,E^*(H_2))\nonumber\\
&=&\max_{H_1} (\bra{\psi}H_1\otimes 1\ket{\psi}-\,E^*(H_1))
+\max_{H_2} (\bra{\psi}1\otimes H_2\ket{\psi}-\,E^*(H_2))
\eea
\end{widetext}
An equivalent statement is that for all $\psi$, $H_1$ and $H_2$
\beq
E(\psi)\coin\bra{\psi}(H_1\otimes 1+1\otimes H_2)\ket{\psi}-\,E^*(H_1)-\,E^*(H_2).
\eeq  
One can further rewrite it as follows
\beq
\bra{\psi}(H_1\otimes 1+1\otimes H_2)\ket{\psi}-\,E(\psi)\colein\,E^*(H_1)+\,E^*(H_2).
\eeq
The inequality above is true for all $\psi$ if and only if it is true for the
maximal value of the l.h.s.:
\beq
\max_{\psi}(\bra{\psi}(H_1\otimes 1+1\otimes H_2)\ket{\psi}-E(\psi))\colein\,E^*(H_1)+\,E^*(H_2),
\eeq
that is \cite{ab}:
\beq\label{dsub}
E^*(H_1\otimes 1+1\otimes H_2)\colein\,E^*(H_1)+\,E^*(H_2).
\eeq
On the other hand, let us consider vectors $\psi_1$ and $\psi_2$, optimal (in the sense
of the definition of the conjugate function Eq. (\ref{ddef})) for $H_1$ and $H_2$ respectively.
Using their product $\ket{\psi_1}\ket{\psi_2}$ as a trial function for finding 
$E^*(H_1\otimes 1+1\otimes H_2)$ we have
\begin{widetext}
\bea\label{dsup}
E^*(H_1\otimes 1+1\otimes H_2)
&\geq& \bra{\psi_2}\bra{\psi_1}(H_1\otimes 1+1\otimes H_2)\ket{\psi_1}\ket{\psi_2}
-E(\ket{\psi_1}\ket{\psi_2}) \nonumber\\
&=&\bra{\psi_1}H_1\ket{\psi_1}-E(\psi_1)+\bra{\psi_2}H_2\ket{\psi_2}-E(\psi_2)
=E^*(H_1)+\,E^*(H_2).
\eea
\end{widetext}
Eqs. (\ref{dsub}) and (\ref{dsup}) taken together allow one to restate
the strong superadditivity conjecture as the following additivity conjecture
for conjugate functions \cite{ab}:
\beq\label{dadd}
E^*(H_1\otimes 1+1\otimes H_2)\conj\,E^*(H_1)+\,E^*(H_2).
\eeq

\section{Properties of the optimal vectors}
Let us consider in a bipartite system $A-B$ a Hermitian operator $H$
and an optimal (in the sense of the definition of $E^*(H)$,  Eq. (\ref{dpur})) 
vector $\widetilde{\psi}$ for it:
\beq\label{dopt}
E^*(H)=\bra{\widetilde{\psi}}H\ket{\widetilde{\psi}}-E(\widetilde{\psi}).
\eeq
Let us denote by $f(\psi)$ the function, the maximum of which is $E^*(H)$:
\beq
f(\psi)=\bra{\psi}H\ket{\psi}-E(\psi).
\eeq
The necessary condition for $f(\psi)$ to have a maximum at the point $\psi$
is the vanishing of its derivatives: $\delta f(\psi)=0$. To compute the
derivatives we need to return to the definition of $E(\psi)$ and rewrite it
more explicitly in the terms of the components $\psi_{ij}$ of the vector 
$\ket{\psi}$, where the first index refers to the subsystem $A$ and the 
second index refers to the subsystem $B$. One can consider $\psi_{ij}$
as components of a matrix $\psi$. In terms of this matrix the definition 
of $E(\psi)$ becomes
\beq
E(\psi)=-\trace(\psi\psi^\dagger\log_2(\psi\psi^\dagger))=-\trace(\rho\log_2(\rho)), \;\;\;
\rho=\psi\psi^\dagger.
\eeq
One has also $\bra{\psi}H\ket{\psi}=\sum_{ijkl}\psi^*_{ij}H_{ij|kl}\psi_{kl}$.
Let us note that because the trace of a product of matrices is invariant
under cyclic permutations we have $\delta \trace(F(\rho))= \trace( F^{\prime}(\rho)\delta \rho)$
for any function of one variable $F(x)$ and its derivative $F^{\prime}(x)$. To prove
this one can use Taylor series expansion of $F(x)$. In our case $F(x)=-x \log_2 x$ and \\
$F^{\prime}(x)=-\log_2 x -1/\ln 2$, which gives
\beq
\delta E(\psi)=-\trace((\log_2\rho +1/\ln 2)\delta \rho)=-\trace(\delta\rho\log_2\rho).
\eeq 
Substituting here $\rho=\psi\psi^{\dagger}$ we obtain
\beq
\delta E(\psi)=-\trace(\psi^{\dagger}\log_2(\psi\psi^{\dagger})\delta\psi+
\log_2(\psi\psi^{\dagger})\psi\delta\psi^{\dagger}).
\eeq
For the variation of $f(\psi)$ we have now
\beq
\delta f(\psi)=\sum_{ijkl}(\delta\psi^*_{ij}H_{ij|kl}\psi_{kl}+
\psi^*_{ij}H_{ij|kl}\delta\psi_{kl})-\delta E(\psi).
\eeq
The vector variation $\ket{\delta\psi}$ is orthogonal to $\ket{\psi}$ due to the normalization 
condition $\braket{\psi}{\psi}=1$, but otherwise arbitrary. Instead of the real and imaginary 
parts of its components $\Re(\delta\psi_{ij})$ and $\Im(\delta\psi_{ij})$, one can consider 
as independent their complex linear combinations $\delta\psi_{ij}$ and $\delta\psi^*_{ij}$.
Then the necessary condition for maximum reads
\beq
\sum_{kl}H_{ij|kl}\widetilde{\psi}_{kl}+
(\log_2(\widetilde{\psi}\widetilde{\psi}^{\dagger})\widetilde{\psi})_{ij}=C \widetilde{\psi}_{ij}.
\eeq
By taking the scalar product of both sides of this equation with $\bra{\widetilde{\psi}}$
we find that $C=E^*(H)$. Taking this into account we finally have
\beq\label{dcop}
\sum_{kl}H_{ij|kl}\widetilde{\psi}_{kl}=
-(\log_2(\widetilde{\psi}\widetilde{\psi}^{\dagger})\widetilde{\psi})_{ij}+E^*(H)\widetilde{\psi}_{ij}.
\eeq
This equation determines how the operator $H$ acts on the optimal vectors
and therefore it determines how it acts on any linear combination of them. 
The Hermiticity of $H$ requires that for any pair of optimal vectors 
$\widetilde{\psi}_\alpha$ and $\widetilde{\psi}_\beta$ the following condition holds:
\beq
\trace\left[\widetilde{\psi}_\alpha \widetilde{\psi}^{\dagger}_\beta 
(\log_2(\widetilde{\psi}_\alpha \widetilde{\psi}^{\dagger}_\alpha)
-\log_2(\widetilde{\psi}_\beta \widetilde{\psi}^{\dagger}_\beta))\right]=0.
\eeq

\section{Connection between the additivity and the strong superadditivity}
The following theorem links the additivity and the strong superadditivity
of the entanglement of formation.
\bigskip

{\bf Theorem:} For an arbitrary state of the whole system (consisting
of 4 parts: $1A$, $1B$, $2A$ and $2B$) with the corresponding density 
matrix $\rho$, let us compute its partially reduced density matrices
$\rho_1=\trace_2(\rho)$ and $\rho_2=\trace_1(\rho)$ . If for these two
density matrices $\rho_1$ and $\rho_2$ the EoF is additive, that is if
\beq\label{ladd}
E_F(\rho_1\otimes\rho_2)=E_F(\rho_1)+E_F(\rho_2)
\eeq
then the EoF is strongly superadditive for the state $\rho$:
\beq\label{goal}
E_F(\rho)\geq E_F(\rho_1)+E_F(\rho_2).
\eeq

\vspace{1mm}
 
{\em Proof:} Let us consider a Hermitian matrix $H$, optimal for $\rho_1 \otimes \rho_2$
in the sense of Eq. (\ref{dd}), that is
\beq\label{l1a}
E_F(\rho_1\otimes\rho_2)=\trace\left[H(\rho_1\otimes\rho_2)\right]-E^*(H).
\eeq
From the definition of the conjugate function (Eqs. (\ref{ddef}) and (\ref{dpur})) we have also:
\beq\label{l1b}
E^*(H)\geq \bra{\psi}H\ket{\psi}-\,E(\psi),\;\;\;E^*(H)\geq \trace(H\rho^{\prime})-\,E_F(\rho^{\prime}),
\eeq
for all pure states $\psi$ and all density matrices $\rho^{\prime}$. Let 
\beq\label{dec1}
\rho_1=\sum_m p^{(1)}_m \proj{\psi^{(1)}_m},\;\;\; p^{(1)}_m> 0,\;\;\; E_F(\rho_1)=\sum_m p^{(1)}_m E(\psi^{(1)}_m)
\eeq
be an optimal decomposition for $\rho_1$ and let
\beq\label{dec2}
\rho_2=\sum_n p^{(2)}_n \proj{\psi^{(2)}_n},\;\;\; p^{(2)}_n> 0,\;\;\; E_F(\rho_2)=\sum_n p^{(2)}_n E(\psi^{(2)}_n)
\eeq
be an optimal decomposition for $\rho_2$. Then for all $m$ and $n$ the products 
$\ket{\psi^{(1)}_m}\ket{\psi^{(2)}_n}$ are optimal pure states for $H$ in the sense of Eq. (\ref{dpur}):
\beq\label{l2}
E^*(H)=\bra{\psi^{(1)}_m}\bra{\psi^{(2)}_n}H\ket{\psi^{(1)}_m}\ket{\psi^{(2)}_n}
-\,E(\ket{\psi^{(1)}_m}\ket{\psi^{(2)}_n}).
\eeq
Indeed, substituting the decompositions Eq. (\ref{dec1}) and Eq. (\ref{dec2}) in 
the Eqs. (\ref{l1a}) and (\ref{l1b}) we have
\bea
E^*(H)&=&\sum_{mn} p^{(1)}_m p^{(2)}_n (\bra{\psi^{(1)}_m}\bra{\psi^{(2)}_n}H\ket{\psi^{(1)}_m}\ket{\psi^{(2)}_n}
-\,E(\ket{\psi^{(1)}_m}\ket{\psi^{(2)}_n})),\nonumber\\
E^*(H)&\geq&\bra{\psi^{(1)}_m}\bra{\psi^{(2)}_n}H\ket{\psi^{(1)}_m}\ket{\psi^{(2)}_n}
-\,E(\ket{\psi^{(1)}_m}\ket{\psi^{(2)}_n}),
\eea
with all probabilities strictly positive:
\beq
p^{(1)}_m p^{(2)}_n> 0, \;\;\;\sum_{mn} p^{(1)}_m p^{(2)}_n = 1.
\eeq
Clearly, this is possible only if Eq. (\ref{l2}) holds for all $m$ and $n$.

Let us denote by $V_1$ the subspace spanned by the vectors $\psi^{(1)}_m$, 
its orthogonal complement by $V_1^{\perp}$, and  by $V_2$ and $V_2^{\perp}$ the 
analogous subspaces for the subsystem $2$. Let us note, that the state $\rho$
must be an ensemble of linear combinations of the optimal optimal vectors 
$\ket{\psi^{(1)}_m}\ket{\psi^{(2)}_n}$, that is an ensemble of pure states 
from $V_1 \otimes V_2$:
\beq
\rho=\sum_{k} p_k \proj{\phi^k}, \;\;\; \phi^k\in V_1 \otimes V_2. 
\eeq
Indeed, we have 
$\phi^k\in \left[(V_1^\perp \otimes V_2) \oplus (V_1^\perp \otimes V_2^\perp)\right]^\perp$,
because for any vector $\ket{v}\in V_1^\perp$ the orthogonality relation 
$\sum_i\phi^{k*}_{ij}v_i=0$ (the first index $i$ corresponds here to the subsystem 
$1$ and the second index $j$ corresponds to the subsystem $2$)  follows from 
$\sum_{jk} |\sum_i\phi^{k*}_{ij}v_i|^2=\bra{v}\rho_1 \ket{v}=0$. Analogously,
one has $\phi^k\in \left[(V_1\otimes V_2^\perp) \oplus (V_1^\perp \otimes V_2^\perp)\right]^\perp$, 
and then
\beq\label{l3}
\phi^k\in \left[(V_1^\perp \otimes V_2) \oplus (V_1\otimes V_2^\perp) \oplus (V_1^\perp \otimes V_2^\perp)\right]^\perp
=V_1 \otimes V_2.
\eeq
Now, let us show that for the matrix $H$ from (\ref{l1a}) one has
\beq\label{l4}
\trace\left[(\rho_1\otimes\rho_2) H\right]=\trace(H\rho).
\eeq
For this purpose one needs to know only the matrix elements of $H$ between 
states from $V_1 \otimes V_2$ (only such matrix elements are present in (\ref{l4})). 
One can find these elements from Eq. (\ref{dcop}),
writing it down for an optimal vector $\ket{\psi^{(1)}_s}\ket{\psi^{(2)}_t}$ 
and taking the scalar product of both sides of the equation with an optimal vector 
$\bra{\psi^{(1)}_m}\bra{\psi^{(2)}_n}$:
\begin{widetext}
\bea\label{elem}
\bra{\psi^{(1)}_m}\bra{\psi^{(2)}_n}H\ket{\psi^{(1)}_s}\ket{\psi^{(2)}_t}
&=&\,-\trace\left[\psi^{(1)}_s\psi^{(1)\dagger}_m\log_2(\psi^{(1)}_s\psi^{(1)\dagger}_s)\right]
\trace(\psi^{(2)}_t\psi^{(2)\dagger}_n)
-\trace\left[\psi^{(2)}_t\psi^{(2)\dagger}_n\log_2(\psi^{(2)}_t\psi^{(2)\dagger}_t)\right]
\trace(\psi^{(1)}_s\psi^{(1)\dagger}_m)\nonumber\\
&+& E^*(H)\trace(\psi^{(1)}_s\psi^{(1)\dagger}_m)\trace(\psi^{(2)}_t\psi^{(2)\dagger}_n).
\eea
\end{widetext}
Here we have written the scalar products as the traces of matrix products,
for example $\braket{\psi^{(1)}_m}{\psi^{(1)}_s}=\trace(\psi^{(1)}_s\psi^{(1)\dagger}_m))$,
and have used the fact that the logarithm of a tensor product of matrices is
the sum of logarithms of the matrices: 
$\log_2(X\otimes Y)=\log_2(X)\otimes 1+ 1\otimes \log_2(Y)$.
We have used also the multiplicativity of trace operation: $\trace(X \otimes Y)=\trace(X)\trace(Y)$.
It is easy to see from Eq. (\ref{elem}), that the matrix elements of $H$ between the
states from $V_1\otimes V_2$ have the form:
\beq
\bra{\psi^\prime}H\ket{\psi}=\bra{\psi^\prime}(H_1\otimes 1 +1\otimes H_2)\ket{\psi},
\eeq 
for some matrices $H_1$ and $H_2$. Then Eq. (\ref{l4}) follows from this formula 
applied to the expectation value $\trace(H\rho)$.

Now, we have all necessary means to prove the theorem statement. Replacing 
$\rho^\prime$ with $\rho$ in the second inequality in Eq. (\ref{l1b}) we obtain the 
following inequality:
\beq
E^*(H)\geq \trace(H\rho)-\,E_F(\rho),
\eeq
If one uses Eqs. (\ref{l1a}) and (\ref{ladd}) to find $E^*(H)$ this inequality takes
the form:
\beq
\trace\left[H(\rho_1\otimes\rho_2)\right]-E_F(\rho_1)-E_F(\rho_2)\geq \trace(H\rho)-\,E_F(\rho).
\eeq
Finally, taking into account Eq. (\ref{l4}) one obtains the inequality (\ref{goal})
which is the statement of the theorem.\qed

The theorem above states that the strong superadditivity of EoF holds for a state $\rho$
if additivity holds for its reduced density matrices $\rho_1$ and $\rho_2$. Then it is clear
that the strong superadditivity of EoF for all states of the system follows from 
the additivity of EoF for all states $\rho_1$ and $\rho_2$ of the subsystems $1$ and $2$.
If the additivity conjecture is not true in general, the above theorem will be
still useful, because it connects the strong superadditivity of a state with
the additivity for its reduced density matrices $\rho_1$ and $\rho_2$ only and does
not require the additivity for all states.

\section{Concluding remarks}
The conjectures stating that the entanglement of formation and the 
Holevo-Schumacher-Westmorland classical capacity of a 
quantum channel are additive have not been proved in general case, but they are 
supported by a number of numerical calculations and they were proved in some 
particular cases. No counterexamples has been found. It was shown in \cite{msw},
that both conjectures are true if EoF has the strong superadditivity property.
The purpose of the present paper was to deepen this connection by establishing
that the strong superadditivity of EoF follows from its additivity and
thus the two conjectures are equivalent. This fact makes more important
the further study of both of them. The strong superadditivity conjecture
which until now seemed rather speculative becomes as plausible as the additivity 
conjecture. It becomes clear that it is not by chance that all known proofs
of additivity of EoF for particular subspaces of quantum states \cite{dvc} are 
based on proofs of the strong superadditivity conjecture for these subspaces,
and finding a counterexample to the former would immediately give a counterexample
for the latter. And the study of the EoF's additivity problem becomes
even more important than it was before, because now its proof would automatically
give a proof of the additivity of the classical channel capacity.

After this work has been completed, the preprint \cite{shor} appeared,
containing among other results a proof of the equivalence of the additivity
and strong superadditivity of the EoF, which is the main result of the present work.
Our proof is analogous but uses a different language and thus can be useful 
for some readers.

\begin{acknowledgments}
I am greatly indebted to Stefano Bettelli for many fruitful discussions
on this problem and for a careful reading of the manuscript and to 
Dima Shepelyansky for stimulating my interest to the problem. 
This work was supported in part by the NSA and ARDA under ARO 
contract No. DAAD19-01-1-0553
\end{acknowledgments}

\end{document}